\documentclass[12pt]{article}
\usepackage{a4}
\usepackage{epsfig}
\begin{document}
\title {Quantum vacuum pressure on a conducting slab}
\author{Marco Scandurra \thanks{scandurr@lns.mit.edu}\\
{\itshape MIT Center for Theoretical Physics}\\
{\itshape and Laboratory for Nuclear Science}\\
{\itshape 77 Massachusetts Avenue, Cambridge, MA 02139}}\maketitle
\begin{abstract}
\noindent 
 The casimir pressure on a non ideal conducting slab is calculated. Using a simple model for the conductivity according to which the slab is perfectly conducting at frequencies below plasma frequency $\omega_p$ and perfectly transparent above such frequency, it is found that the vacuum pressure on each surface of the slab  is $\frac{\hbar\omega_p^4}{24 \pi^2 c^3}$ which is finite without removal of any divergence.
\end{abstract}
\vspace*{-\bigskipamount}

\maketitle

\section{Introduction}
Electromagnetic vacuum fluctuations exert forces on macroscopic objects; the most relevant example of this is the Casimir effect, the attraction between parallel metallic plates in vacuum  \cite{Casimir}. The Casimir force can be envisioned as resulting from scattering of electromagnetic modes by the plates. Because of a  different distribution of the modes in the two regions: between the plates and outside the plates,  a difference in radiation pressure arises which pushes the boundaries together, the force depends on the  fourth power of the plates separation. Casimir experiments \cite{Lamoreaux,Mohideen,Carugno,Capasso} are a direct way of measuring quantum vacuum pressure on macroscopic surfaces, however the configuration of two  parallel plates is not the most elementary one, the pressure is also present on every isolated rigid body made of conducting material. If the body is large enough every element of its surface can be considered as flat and in this case there is no geometrical parameter controlling the vacuum pressure \footnote{Similarly the pressure from blackbody radiation on a large surface is simply given by one third of the energy density} . In this paper I will investigate the radiation pressure on a thick metallic slab. Although the resulting force will turn out to be independent of the thickness of the slab, it will strongly depend on the electrodynamic properties of the material the slab is made of. In Casimir theory metals are frequently studied as perfect conductors. If a metallic slab is assumed to be a perfect conductor the vacuum pressure on it is infinite. However the assumption of perfect conductivity is unphysical, a finite pressure can be found studying a realistic conductor whose interaction with the electromagnetic field is governed by the plasma frequency, this is the program of the present article. Calculations of vacuum energies for compact objects, like spheres and cylinders were undertaken by several authors in the past, in particular dielectric spheres and cylinders  attracted the attentions of investigators \cite{Barton,Bordag,Brevik}. Removal of divergences was accomplished by these authors using several elegant techniques, including  zeta functional renormalization. In the present paper no subtraction of terms is needed, the plasma frequency, which plays the role of a coupling constant regulates the energy. Metals behave as excellent conductors for waves of frequency $ \omega\leq \omega_p $ and as good insulators for frequencies above $\omega_p $, this property is used here to compute the pressure without facing the ultraviolet catastrophe and without removing terms ad hoc. Mode summation is performed  assuming that only modes below plasma frequency contribute to the pressure, while modes with higher frequency propagate in space without suffering  scattering by the slab, this  is explained and justified in detail in the next section. In section III the calculation of the pressure is performed. In section IV the result is briefly discussed.

\section{Plasma theory of conductors}
In this section the classical electrodynamics of wave propagation inside metals is summarized. The wave vector in a conductor is given by
\begin{equation}
k^2=\frac{\omega^2}{c^2}\left( 1-i\frac{\sigma (\omega)}{\omega \epsilon_0}\right)
\end{equation}
where $\sigma(\omega)$ is the conductivity and $\epsilon_0$ is the permittivity of vacuum.  According to Drude's model the conductivity is given by
\begin{equation}
\sigma(\omega)\ =\ \frac{N_e e^2}{m_e}\left(\frac{1}{\nu+i\omega}\right)
\end{equation}
where $N_e$ is the density of free electrons, $m_e$ is the mass of the electron and $\nu$ is the frequency of collision with the lattice atoms, which for good conductors like silver is of the order of $10^{13}$ Hz. If the frequency $\omega$ of a wave entering a metal is lower than the collision frequency, the imaginary term in (2) can be neglected and  we are in the normal skin-effect regime, i.e. the fields possess a propagation term and a damping term governed by the length $\sqrt{2\epsilon_0 c^2/(\omega \sigma)}$. On the contrary, when $\omega>>\nu$, damping collisions with the atoms become insignificant and the electrons behave like a plasma. In most of Casimir experiments this is the relevant regime and we will deal with it for the remainder of this article. The wave vector becomes
\begin{equation}
k=\pm \frac{1}{c}\sqrt{\omega^2-\omega_p},
\end{equation}
where $\omega_p=\sqrt{N_e e^2/(m_e\epsilon_0)}$. The electric field penetrating inside the metal along the direction $z$ is
\begin{equation}
E=E_0\left(e^{-\frac{z}{c}\sqrt{\omega_p^2-\omega^2}}\  e^{-i\omega t}\right)\ .
\end{equation}
We see that (4) contains no wave term as long as $\omega <\omega_p$. This means that a wave below plasma frequency is exponentially  damped inside the metal, but unlike the normal skin-effect  damping, there is no oscillatory term and the attenuation of the field is more rapid. When $\omega$ approaches $\omega_p$ the metals becomes an ideal conductor and the electromagnetic field inside the metal vanishes. Finally when $\omega >\omega_p$ a wave term appears and the wave can propagate as in a dielectric. Given this kind of behavior one is led almost naturally to consider the metal as an ideal conductor for frequencies below plasma frequency and as an ideal transparent medium (n=1) for higher frequencies. And assumed that in traversing a perfectly transparent medium waves do not loose energy or momentum, we can neglect them in the computation of Casimir forces and perform mode summation only up to the highest integer number corresponding to plasma frequency. This is of course a rough approximation, the advantages of which  rest on the possibility of analytical calculations with qualitatively correct results.

\section{Pressure on a flat surface}
 Let us consider a plane solid slab made of a metal or a conducting material with plasma frequency $\omega_p$. The slab has a finite thickness $a$ along the $z$-axis and extends infinitely in the other directions $x$ and $y$.  The slab is embedded by the electromagnetic vacuum. By enclosing space in a cubic quantization box of size $L$ one has the vacuum energy
\begin{equation}
E_0\ =\ \frac 12 \sum_{(n)}^\infty\ \hbar \omega_{(n)}\ ,
\end{equation}
where 
\begin{equation}
\omega_{(n)}=\hbar c \sqrt{(\pi l/L)^2 + (\pi m/L)^2 + (\pi n/L)^2}\ .
\end{equation}
In eq.(5) ${(n)}$ includes  all possible quantum numbers and $l,m,n$ are positive integers\footnote{Boundary conditions are assumed temporarily on the quantization box}. Quantity (5) is divergent, however we will only sum up to an integer corresponding to $\omega_p$ for the reasons already explained, higher frequency modes are ignored; the result will be finite without renormalization. The slab is considered as  perfectly impenetrable  for low frequency modes. Each mode yields a momentum $2\hbar K_z$ upon scattering, $K_z$ being the component of the wave vector normal to the slab.  None of such modes lives in the interior of the slab and there is no internal (outward) pressure, thus the interior of the slab is completely ignored. The contribution to the radiation pressure coming from a single low frequency mode with wave vector $k=\sqrt{(\pi l/L)^2 + (\pi m/L)^2 + (\pi n/L)/^2}$ is
\begin{equation} 
P_{k,\epsilon}= \frac {1}{2 L^3} 2 \hbar k c \cos^2\theta 
\end{equation}
where $\theta$ is the angle between the wave vector and the normal above the surface. Summation over $l,m,n$ and over the two polarization states $\epsilon$ gives the total pressure. Since the space on both sides of the slab is unbounded a triple integral over the components $K_x,K_y,K_z$  can replace the summation. The limits of integration are chosen so that the maximum value of $k=\sqrt{K_x^2+K_y^2+K_z^2}$ is $\omega_p/c$. Keeping in mind that $\cos\theta=K_x/k$ and the fact that only a half of the modes are moving toward the boundary, the total pressure is :
\begin{equation} 
P = \hbar c \int_0^{\pi/2}d\phi\int_0^{\omega_p/c}dK_z \int_0^{\sqrt{\omega_p^2/c^2-K_z^2}} dq q \frac{K_z^2}{\sqrt{q^2 + K_z^2}} 
\end{equation}
where $q=\sqrt{K_x^2+K_y^2}$ and $\phi$ is the angle in the $x-y$ plane.
After carrying out all integrations one finds:
\begin{equation}
P=\frac{\hbar \omega_p^4}{24 \pi^2 c^3} 
\end{equation}
This result, which is independent of the slab thickness, represents the force that the vacuum exerts per unit surface of the conducting slab, it is directed toward the interior of the slab.

\section{Discussion}
In this article the pressure of the electromagnetic quantum vacuum on a conducting slab was calculated. The pressure is found to depend on the fourth power of the plasma frequency of the metal and  therefore it depends on the square of the carrier concentration in the material. If the slab is made of silver, $\omega_p=1.36\  10^{16}$ rad/sec and the vacuum pressure according to (9) is about 500 Newton per square meter, which is significant. Formula (9) is in principle valid for all kind of geometries, including metallic balls and cylinders, as long as the object is large compared to the plasma wavelength. When the object is small a different treatment is necessary, for instance in the case of a tiny compact sphere, the vacuum pressure can be calculated my means of Raleigh scattering formulas while  assumptions made in section II will still hold.  Eq (9) is valid only when the thickness of the slab is larger than the penetration depth of the relevant waves, i.e. $a$ must be larger than the plasma wavelength $\lambda_p=2\pi c/\omega_p$. The present article neglects the physics of the interior of the metal including interatomic forces of the lattice. Although the experimental implications  are still not clear, result (9) deserves some attention in itself since it relates in a simple way a Casimir force to the physical properties of the boundary. In a debate on the correct way of calculating Casimir energies \footnote{The debate was at the center of the Workshop on Casimir forces held at ITAMP on November 2002}, the present investigation supports the idea that the knowledge of the details of the interaction between the boundary and the quantum field is necessary to calculate Casimir forces consistently. Formula (9) will always give a finite pressure, as long as the plasma frequency of the metal takes physical values.

\vspace{1cm}
\noindent {\bf \Large Acknowledgments}\\

\noindent This work is supported in part by funds provided by the U.S. Department of Energy (D.O.E.) under cooperative research agreement DF-FC02-94ER40818

\end{document}